\documentclass[11pt]{article}

\usepackage[utf8]{inputenc}
\usepackage[T1]{fontenc}
\usepackage[margin=1in]{geometry}
\usepackage{amsmath,amssymb}
\usepackage{booktabs}
\usepackage{graphicx}
\usepackage[authoryear,round]{natbib}
\usepackage[colorlinks=true,allcolors=black,urlcolor=blue]{hyperref}

\title{Assessing Distributional Covariate Balance After Weighting: \\
A Practical Guide}

\author{Ariel Linden\thanks{Corresponding author: Ariel Linden, DrPH, Department of Medicine, Division of Clinical Informatics \& Digital Transformation (DoC-IT), University of California, San Francisco, San Francisco, CA, USA. Email: ariel.linden@ucsf.edu}\\[0.5ex]
Department of Medicine, Division of Clinical Informatics \& Digital Transformation (DoC-IT)\\
University of California, San Francisco, California, USA}

\date{}

\begin{document}

\maketitle

\begin{abstract}
\noindent Health researchers commonly use weighting methods to improve covariate balance when estimating program and intervention effects from observational data. Standardized mean differences (SMDs) and variance ratios assess particular features of balance but may overlook residual differences elsewhere in the covariate distribution. This paper provides practical guidance for using Kolmogorov--Smirnov (KS), Anderson--Darling (AD), and Cram\'er--von Mises (CVM) statistics as complementary distributional balance diagnostics. In an empirical evaluation of a disease management program, weighting based on a main-effects propensity-score model produced SMD and KS statistics below commonly used or proposed balance thresholds of 0.10 for two continuous covariates, which would conventionally be interpreted as indicating acceptable balance. Nevertheless, permutation tests identified residual distributional differences. Adding quadratic terms for these covariates approximately halved their KS statistics, improved variance balance, and eliminated evidence of residual distributional differences, while leaving the estimated program effect substantively unchanged. Supporting simulations showed that KS, AD, and CVM differed in sensitivity to different forms of distributional discrepancy and that the calibration of resampling-based \textit{p}-values depended on how the reference distribution was constructed. These findings support a practical approach that combines conventional balance measures, graphical assessment, complementary distributional diagnostics, weighting-model refinement, and outcome adjustment when important residual imbalance remains.
\end{abstract}


\section{Introduction}\label{sec:intro}

Balance assessment is an integral component of causal inference \citep{rosenbaumrubin1983}. In health professions research, where investigators frequently evaluate programs, treatments, and health care interventions using observational data, demonstrating that treated and comparison groups are comparable on observed baseline characteristics is an important step in supporting a causal interpretation of an estimated treatment effect. Whether covariate balance is achieved by design, as in randomized experiments, or through adjustment, as with propensity-score weighting, matching, or stratification, balance on measured covariates provides an empirical assessment of whether the design or adjustment procedure achieved its intended effect. It does not establish balance on unmeasured characteristics or rule out unmeasured confounding.

Standardized mean differences (SMDs) and variance ratios are commonly used to assess differences in covariate location and dispersion, respectively \citep{austin2009,lindensamuels2013,AustinStuart2015}. An absolute SMD below 0.10 is widely used as a criterion for acceptable mean balance \citep{Normand2001}, while \citet{Rubin2001} proposed variance ratios between 0.5 and 2.0, with a value of 1 indicating equal variances. These measures are useful summaries, but neither characterizes the complete covariate distribution.

Distributional diagnostics provide additional information. The Kolmogorov--Smirnov (KS) statistic measures the largest separation between two empirical cumulative distribution functions \citep{kolmogorov1933,Smirnov1939}, and \citet{Markoulidakis2021} provided simulation-based support for 0.10 as a practical KS balance criterion. The Cram\'er--von Mises (CVM) statistic accumulates discrepancies across the distribution \citep{cramer1928,vonmises1931}, whereas the Anderson--Darling (AD) statistic gives greater relative emphasis to discrepancies toward the tails \citep{anderson1952}. Weighted versions of all three statistics are available \citep{Trusina2020}, but their different sensitivities raise a practical question: how should distributional balance be evaluated when the diagnostics do not provide the same information?

A related issue concerns the interpretation of resampling-based \textit{p}-values after weighting. When weights are estimated from treatment assignment, the resulting reference distribution can depend on how treatment labels and estimated weights are handled during resampling. Thus, both the choice of distributional statistic and the construction of its reference distribution can affect the evidence available to an evaluator.

This paper addresses these issues from a practical perspective. An empirical evaluation of a disease management program for patients with congestive heart failure demonstrates how distributional diagnostics can identify opportunities to improve a weighting specification that satisfies commonly used balance criteria. Supporting simulations examine the relative sensitivity of KS, AD, and CVM to different forms of distributional discrepancy and the behavior of alternative resampling procedures with estimated weights. Together, these analyses inform a workflow for assessing distributional balance and responding to residual imbalance in applied health research. Section~\ref{sec:assess} describes the balance diagnostics and resampling procedures; Section~\ref{sec:example} presents the empirical example; Section~\ref{sec:simulations} summarizes the simulation evidence; Section~\ref{sec:discussion} provides practical guidance and discusses limitations; and Section~\ref{sec:conclusion} concludes.

\section{Assessing Covariate Balance After Weighting}\label{sec:assess}

\subsection{Conventional Balance Diagnostics}\label{sec:conventional}

Covariate balance is commonly assessed using both graphical and numerical diagnostics. Graphical displays can reveal differences across the distribution of a covariate that may not be apparent from numerical summaries alone, although their interpretation is necessarily subjective \citep{lindengraphic2015}. Numerical diagnostics provide reproducible summaries that can be compared across covariates and alternative weighting solutions \citep{lindensamuels2013}.

For a continuous covariate $X$, the absolute standardized mean difference (SMD) between treated ($T$) and comparison ($C$) groups is
\begin{equation}\label{eq:smd}
\mathrm{SMD}=
\frac{|\bar X_T-\bar X_C|}
{\sqrt{(S_T^2+S_C^2)/2}},
\end{equation}
where $\bar X_T$ and $\bar X_C$ are the group means and $S_T^2$ and $S_C^2$ are the corresponding variances. The SMD is dimensionless, allowing balance to be compared across variables measured on different scales. An absolute SMD below 0.10 is commonly interpreted as acceptable balance \citep{Normand2001}, although this threshold is a convention rather than an empirically established boundary between balanced and imbalanced covariates \citep{lindensamuels2013}.

The variance ratio (VR) assesses a different feature of balance,
\begin{equation}\label{eq:vr}
\mathrm{VR}=\frac{S_T^2}{S_C^2}.
\end{equation}
A value of 1 indicates equal variances. \citet{Rubin2001} proposed that values below 0.5 or above 2.0 indicate imbalance. When weighting is used, weighted means and variances replace their unweighted counterparts in Equations~\eqref{eq:smd} and~\eqref{eq:vr} \citep{lindensamuels2013}.

SMD and VR therefore provide complementary information about location and dispersion, but neither characterizes the complete covariate distribution.

\subsection{Distributional Balance Diagnostics}\label{sec:distributional}

A natural starting point for assessing the entire distribution is the weighted empirical cumulative distribution function (ECDF). For group $g$, let $X_{gi}$ denote the covariate value for observation $i$ and $w_{gi}>0$ its weight. The weighted ECDF is
\begin{equation}\label{eq:wecdf}
\widehat F_{gw}(x)=
\frac{\sum_{i=1}^{n_g}w_{gi}I(X_{gi}\le x)}
{\sum_{i=1}^{n_g}w_{gi}}.
\end{equation}

Plotting the weighted ECDFs for the treated and comparison groups shows both the magnitude and location of residual differences. Numerical distributional statistics summarize these differences in complementary ways.

The weighted Kolmogorov--Smirnov statistic is
\begin{equation}\label{eq:ks}
D_w=
\sup_x
\left|
\widehat F_{Tw}(x)-\widehat F_{Cw}(x)
\right|.
\end{equation}

Its magnitude is particularly intuitive: $D_w=0.07$, for example, indicates a maximum separation of seven percentage points between the weighted cumulative distributions. \citet{Markoulidakis2021} provided simulation-based support for $D_w<0.10$ as a practical balance criterion. However, KS retains only the largest separation and does not describe discrepancies occurring elsewhere in the distribution.

The Cram\'er--von Mises (CVM) statistic instead accumulates squared differences between the two ECDFs across the distribution, whereas the Anderson--Darling (AD) statistic gives greater relative emphasis to differences toward the tails. Weighted forms of KS, AD, and CVM have been developed \citep{Trusina2020}. The three statistics consequently need not provide the same assessment when residual imbalance is localized, diffuse, or concentrated in the tails.

Although the magnitude of the KS statistic is directly interpretable, AD and CVM do not have similarly interpretable scales, and none of the three statistics has an empirically established cutpoint that can be assumed to define acceptable covariate balance across settings. Moreover, for weighted versions of these statistics, an appropriate analytic reference distribution is generally not available. Permutation is therefore used to evaluate the observed statistics by comparing them with the distribution of statistics obtained after repeatedly reassigning treatment labels. The resulting permutation \textit{p}-value describes how unusual the observed statistic is under the specified permutation procedure.

\subsection{Resampling Procedures With Estimated Weights}\label{sec:estimated}

An additional issue arises when the weights are estimated from treatment assignment, such as with propensity-score weighting. The observed weights are partly determined by the same treatment labels defining the groups whose distributions are being compared. Construction of a reference distribution must therefore account for what happens to the treatment labels and weights during resampling.

One approach is to hold the estimated weights fixed and permute the treatment labels. Another is to re-estimate the weighting model after each permutation. A third is to generate treatment assignments from an estimated assignment mechanism and then construct the corresponding weights from the fitted assignment probabilities. These approaches represent different resampling procedures and need not produce equivalent \textit{p}-values. Thus, a distributional test that accommodates weights computationally does not necessarily provide a well-calibrated \textit{p}-value for every way in which its reference distribution might be constructed.

\section{Empirical Example}\label{sec:example}

\subsection{Data and Program}\label{sec:data}

The empirical example uses data from a disease management program for patients with congestive heart failure implemented in a large health plan in the Western United States and described previously \citep{linden2014}. Eligible patients were contacted by telephone and invited to enroll. Participants received nurse-directed telephone outreach or remote telemonitoring, as assigned by a program nurse, with the goal of reducing avoidable hospitalizations. Retrospectively collected data include 12 months before and 12 months during the intervention for program participants who completed the 12-month intervention and for non-participants who were health plan members during the same period. Participation is treated as a binary exposure regardless of intervention mode. The outcome is the pre--post difference in all-cause hospital admissions per patient (intervention period minus baseline), so that negative values indicate a reduction in admissions.

The propensity-score model included sex, age, the Charlson Comorbidity Index (CCI) score \citep{Charlson1987}, indicators for complicated diabetes, uncomplicated diabetes, and chronic obstructive pulmonary disease, and five baseline measures of health care utilization: prescriptions filled, office visits, emergency department visits, hospital admissions, and hospital days. Also included was a continuous predictive score estimating each patient's future total health care costs from prior pharmacy claims data \citep{Powers2005}, referred to here as the Rx score. Because this score cannot be calculated for patients with no prescriptions filled during the baseline year, the analytic sample was restricted to patients with a defined Rx score. The resulting sample included 5,907 patients: 766 program participants and 5,141 non-participants. Distributional balance is examined for the Rx and CCI scores to illustrate the diagnostic approach.

\subsection{Weighting and Initial Balance Assessment}\label{sec:weighting}

Propensity scores were first estimated by logistic regression of program participation on the 12 covariates entered as main effects. Stabilized inverse probability of treatment weights targeting the average treatment effect were then calculated \citep{robins2000}. Participants received $\Pr(T=1)/\hat{e}_i$ and non-participants received $\Pr(T=0)/(1-\hat{e}_i)$, where $\hat{e}_i$ is the estimated propensity score for patient $i$. No weights were trimmed or truncated.

Balance was assessed using the SMD, VR, and KS diagnostics described in Section~\ref{sec:assess}, together with weighted KS, AD, and CVM permutation tests computed with the Stata commands \texttt{kstest}, \texttt{adtest}, and \texttt{cvmtest} \citep{lindenadtest2025,lindencvmtest2025,lindenkstest2026}. Permutation \textit{p}-values were based on 2,000 permutations with the estimated weights held fixed. All unweighted and weighted comparisons used the same 5,907 patients.

Table~\ref{tab:balance} summarizes the results. Before weighting, both covariates exceeded the 0.10 SMD and KS criteria. After weighting with the main-effects propensity-score model, the SMDs fell to 0.046 for the Rx score and 0.034 for CCI, while the corresponding KS statistics fell to 0.0706 and 0.0579. The VRs were 0.807 and 0.801. Thus, all three measures fell within their conventional or proposed ranges for acceptable balance.

\begin{figure}[p]
\centering
\includegraphics[width=0.49\textwidth]{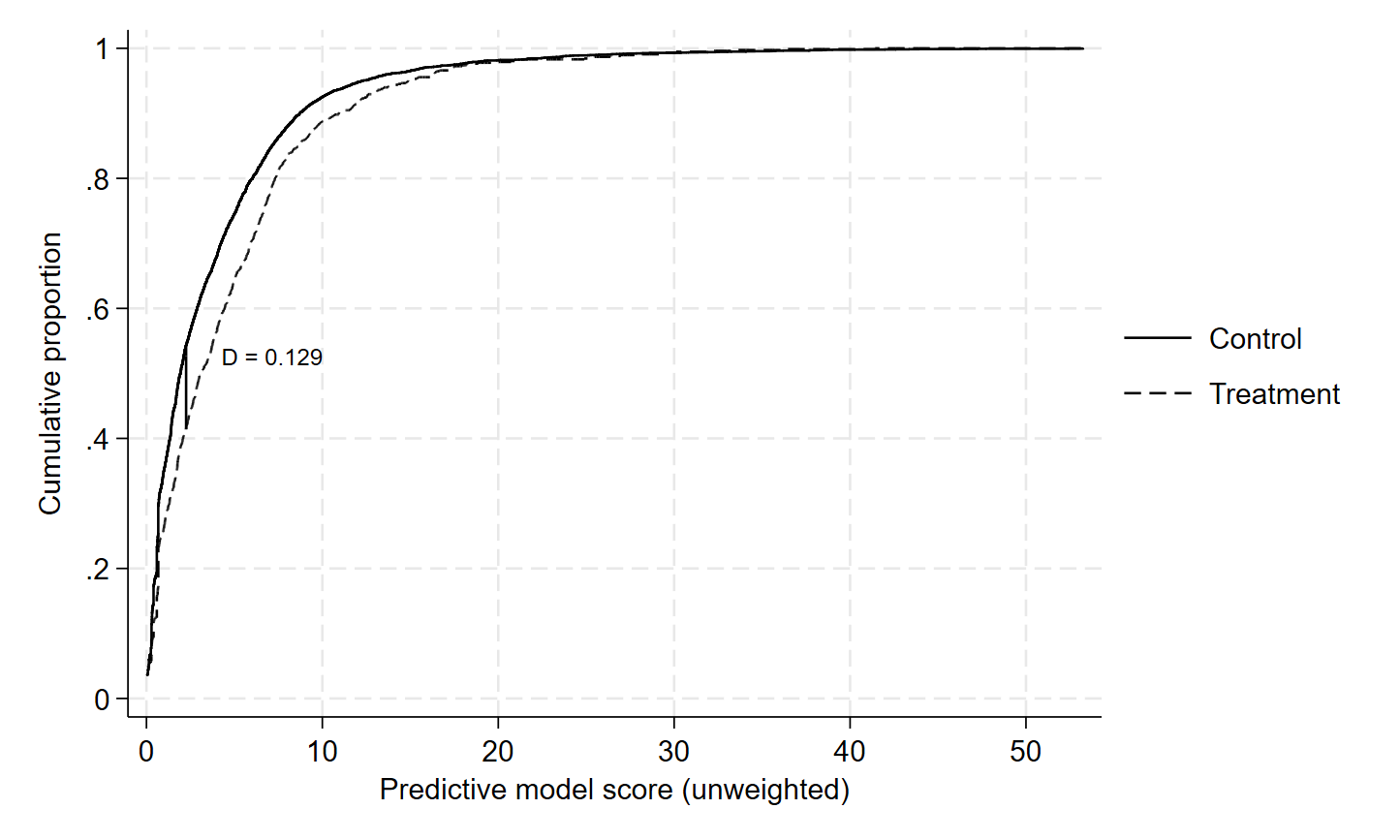}\hfill
\includegraphics[width=0.49\textwidth]{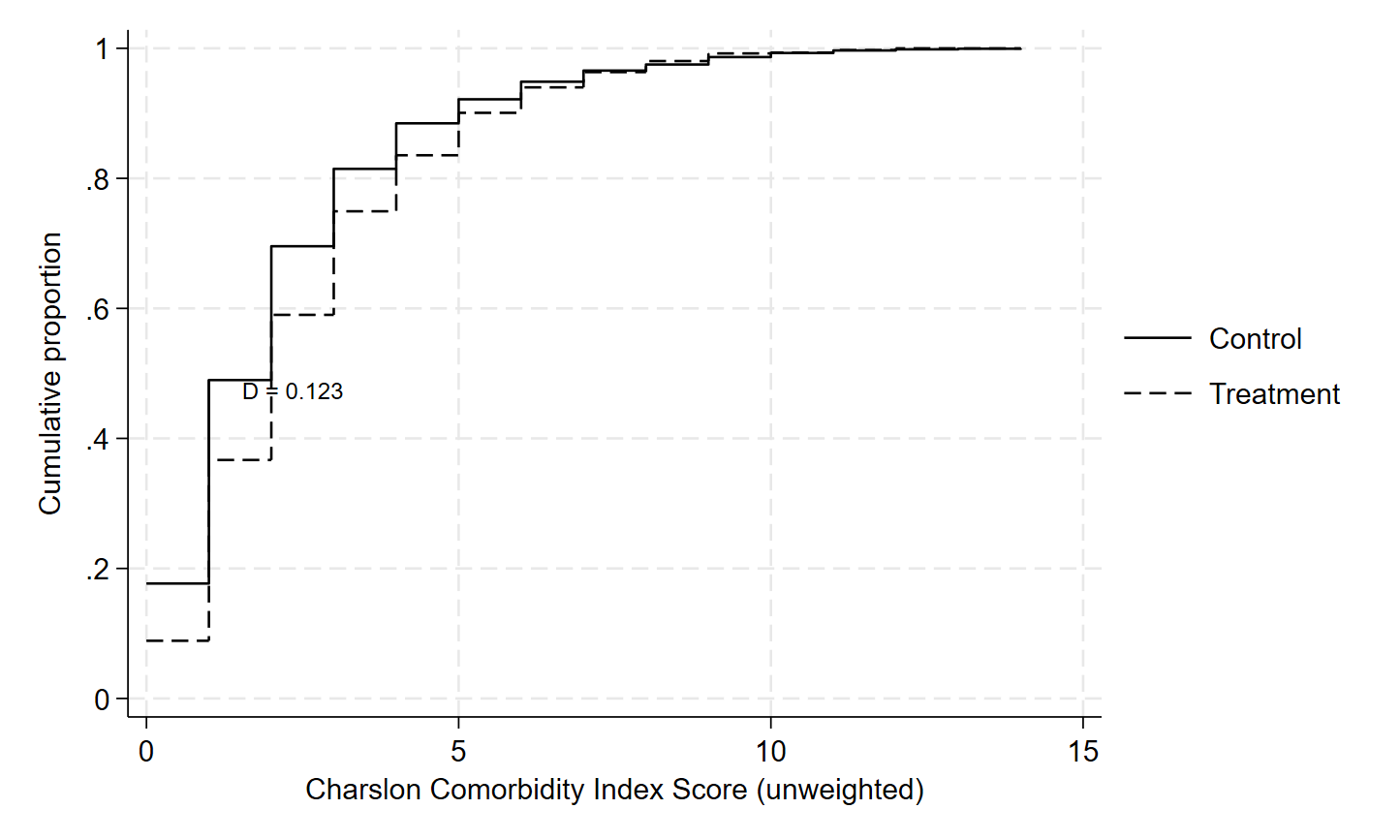}\\[1.5ex]
\includegraphics[width=0.49\textwidth]{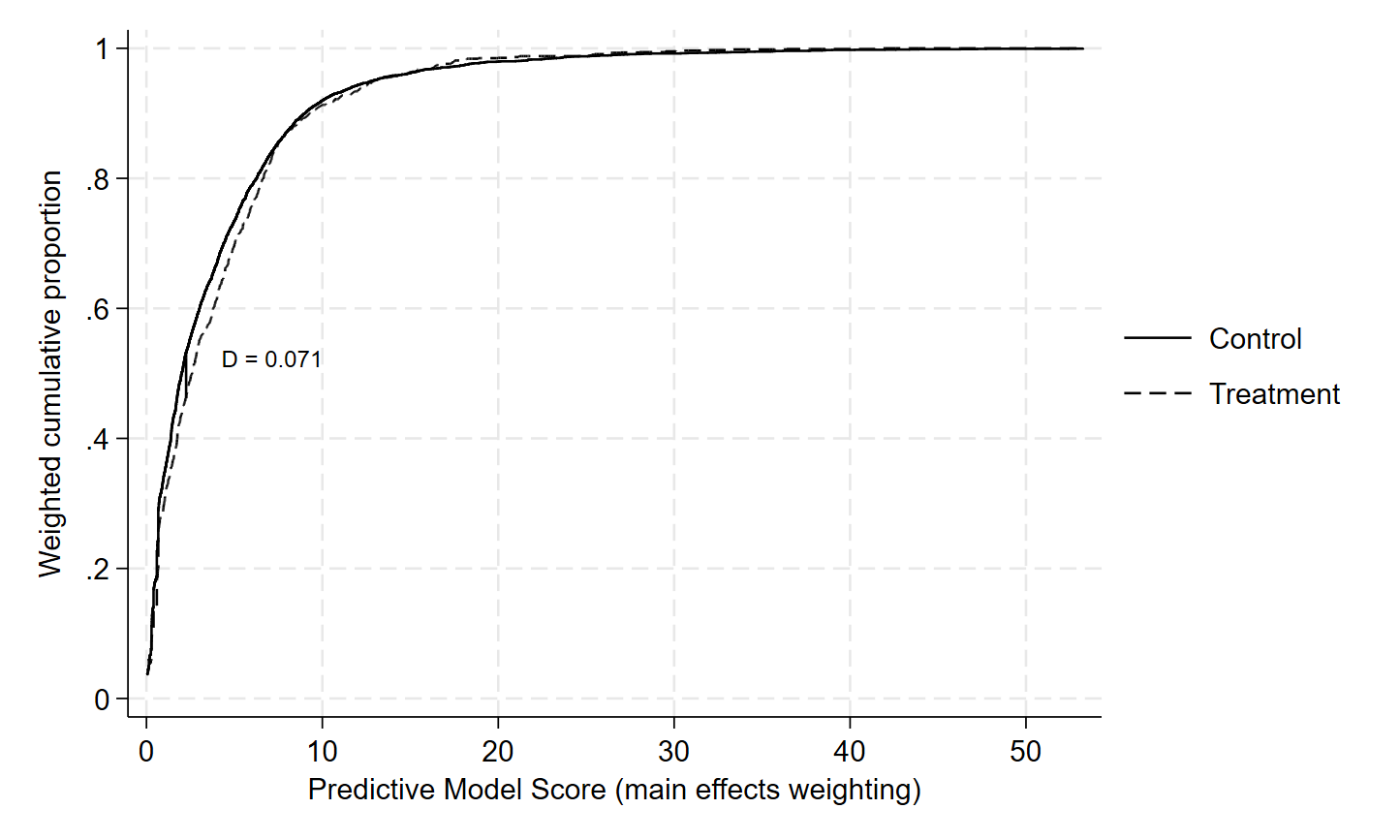}\hfill
\includegraphics[width=0.49\textwidth]{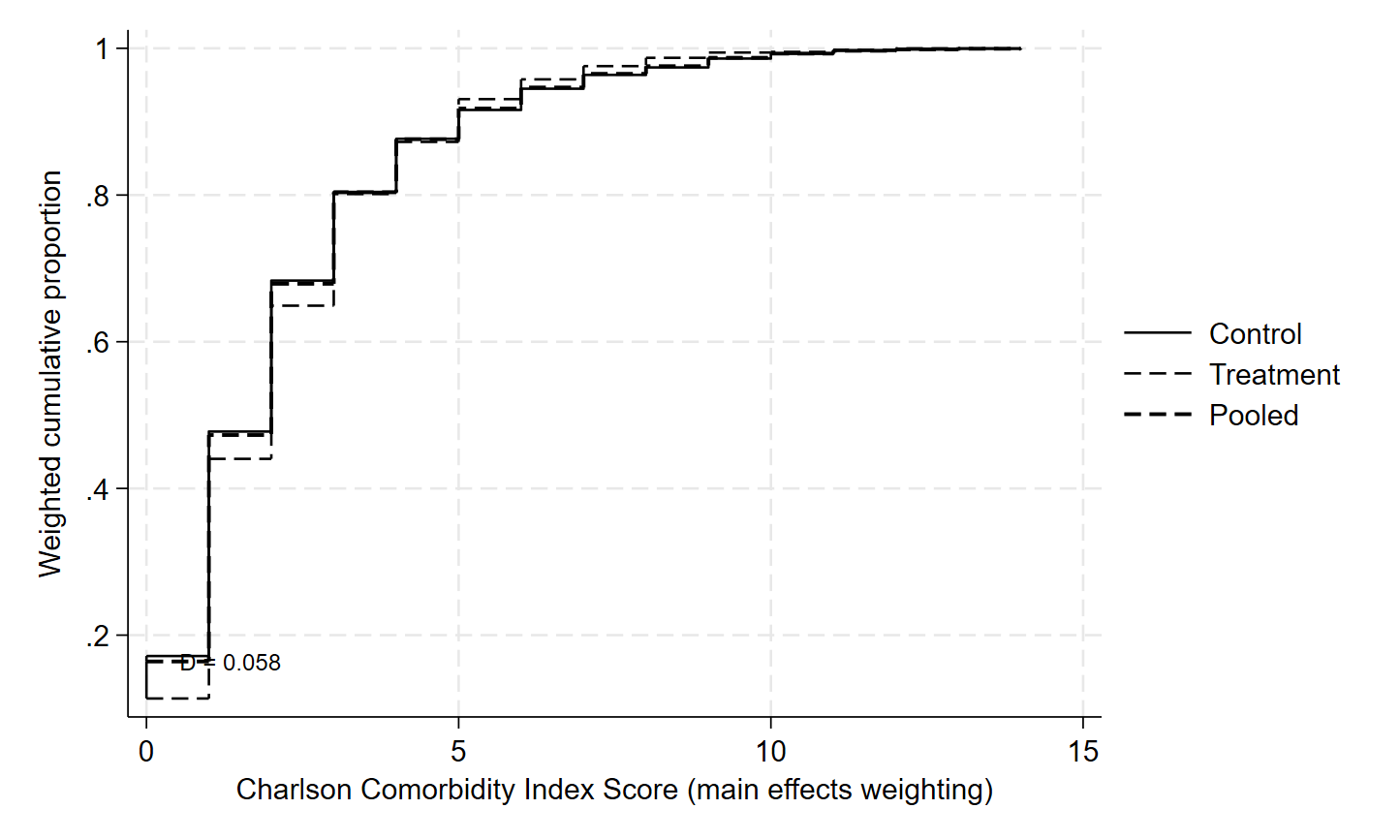}\\[1.5ex]
\includegraphics[width=0.49\textwidth]{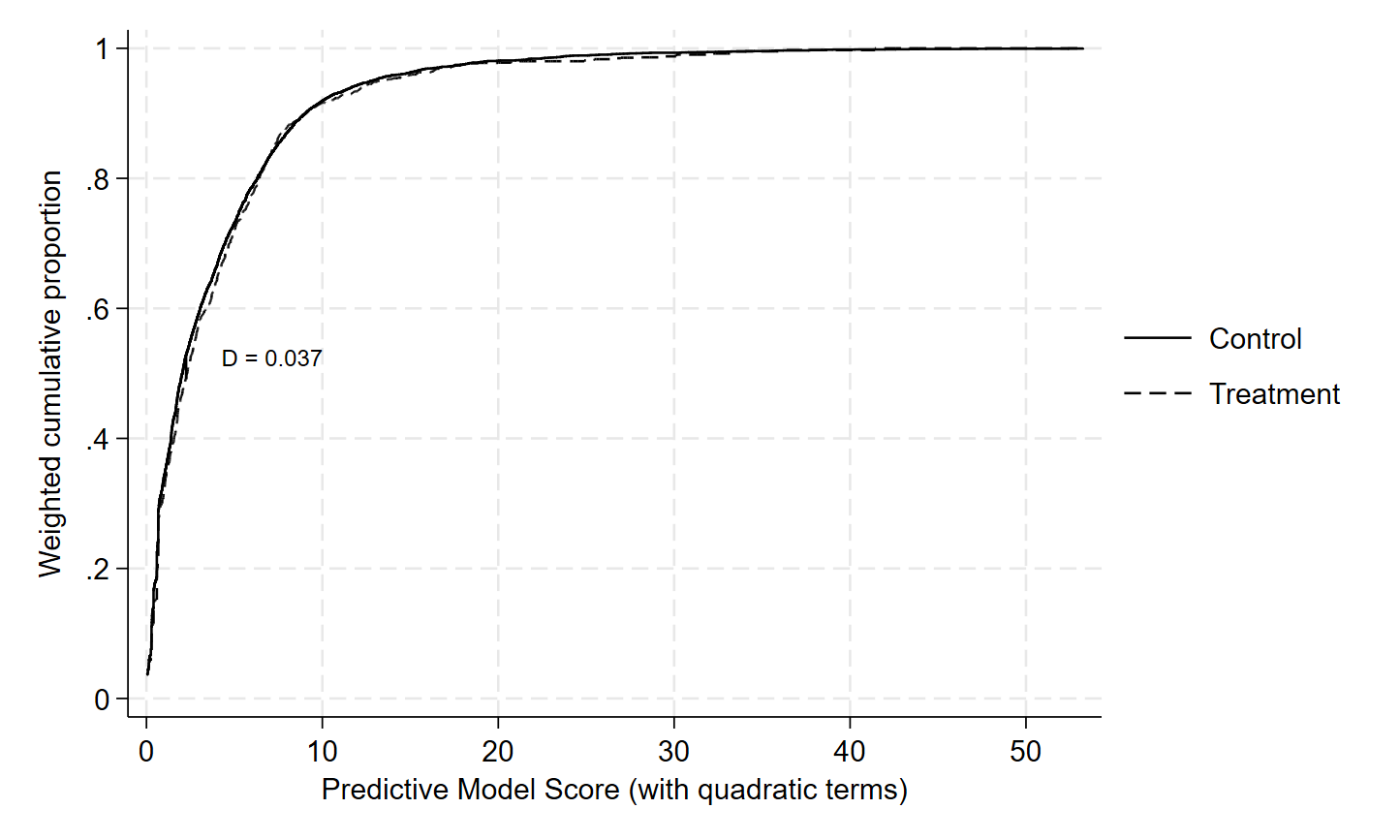}\hfill
\includegraphics[width=0.49\textwidth]{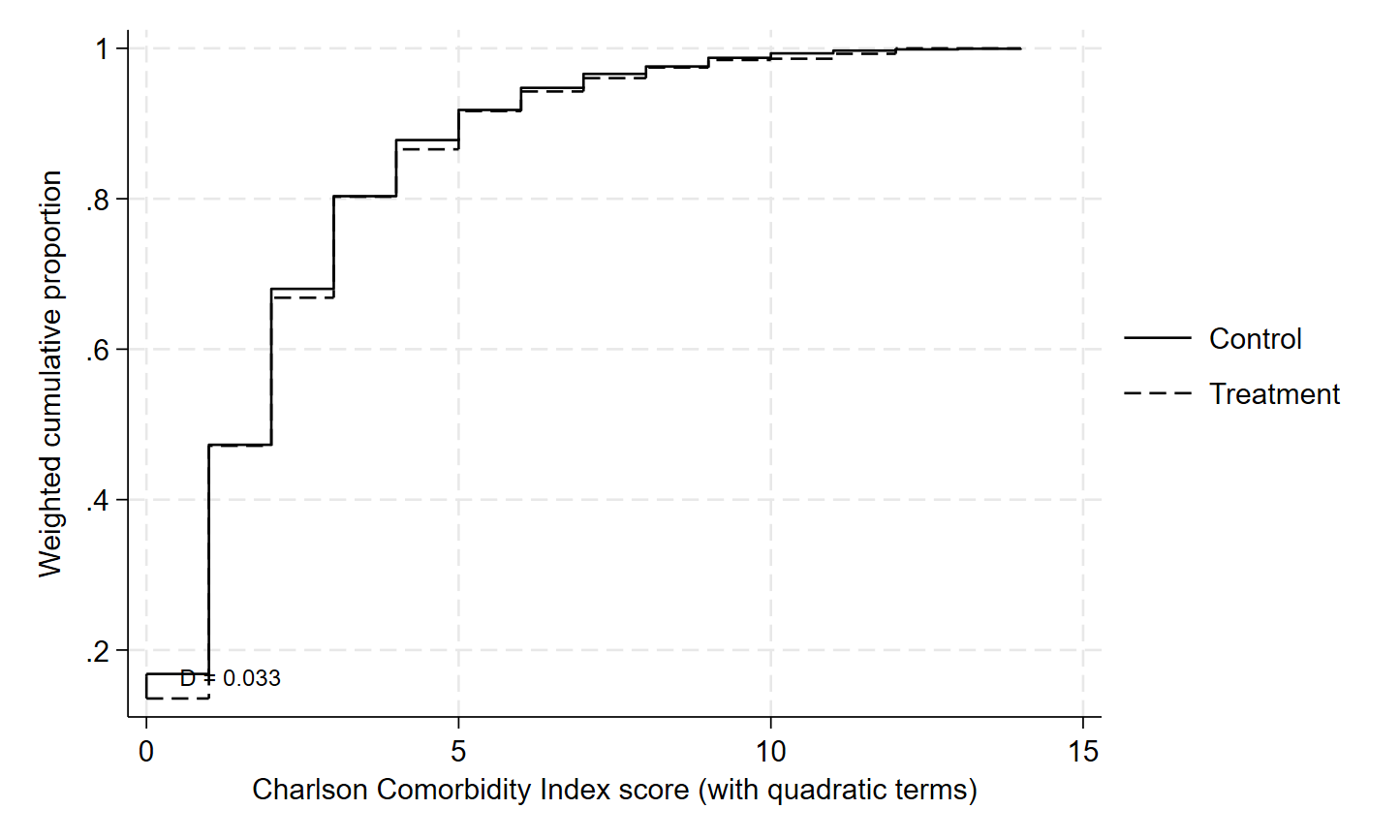}
\caption{Weighted Empirical Cumulative Distribution Functions of the Rx Score (Left) and CCI Score (Right) by Program Participation: Unweighted (Top), Main-Effects Weights (Middle), and Weights With Quadratic Terms (Bottom)}
\label{fig:ecdf}
\par\vspace{1ex}
{\footnotesize \textit{Note.} \textit{D} = Kolmogorov--Smirnov statistic (the maximum vertical distance between the two group curves). ``Treatment'' denotes program participants and ``Control'' denotes non-participants.\par}
\end{figure}

The permutation tests provided a different perspective. KS, AD, and CVM all indicated residual distributional differences for both covariates, with \textit{p}-values ranging from .002 to .009. The weighted ECDFs in Figure~\ref{fig:ecdf} show where these remaining differences occurred. Thus, weighting had clearly improved balance, but satisfying the SMD, VR, and proposed KS criteria did not imply that the weighted distributions were indistinguishable.

\subsection{Refining the Propensity-Score Specification}\label{sec:refine}

Both continuous covariates had been included in the original propensity-score model as linear main effects. To determine whether a more flexible functional form could improve their distributional balance, quadratic terms for the Rx and CCI scores were added while all other predictors were left unchanged. Propensity scores and stabilized weights were then re-estimated, and balance was reassessed for the original covariates.

The refinement produced little change in mean balance: the SMD changed from 0.046 to 0.041 for the Rx score and from 0.034 to 0.039 for CCI. Other aspects of balance improved substantially. The VRs moved from 0.807 to 1.096 and from 0.801 to 1.060, respectively, while the KS statistics declined from 0.0706 to 0.0375 for the Rx score and from 0.0579 to 0.0326 for CCI. None of the six permutation tests provided evidence of residual distributional differences under the refined weights (\textit{p}-values .170--.400).

Both weighting specifications satisfied the conventional SMD and VR criteria and the proposed KS criterion of 0.10, yet the refined specification produced markedly smaller distributional discrepancies. Thus, treating these criteria as stopping rules would have obscured an achievable improvement in balance.

\subsection{Treatment-Effect Estimates}\label{sec:effects}

The final question was whether improving distributional balance materially changed the estimated program effect. The pre--post change in admissions was regressed on program participation using each set of stabilized weights and robust standard errors; an unweighted analysis of the same 5,907 patients is included for comparison. Table~\ref{tab:effects} reports the group-specific mean changes and treatment-control contrasts.

With main-effects weights, admissions decreased by an estimated 0.229 admissions per patient among participants and 0.341 among non-participants, yielding a difference of 0.112 admissions (95\% CI, 0.010--0.215). With the refined weights, the corresponding decreases were 0.218 and 0.342, yielding a difference of 0.124 (95\% CI, 0.023--0.224). Thus, under both weighting specifications, non-participants experienced a greater reduction in admissions than program participants. Refining the propensity-score model substantially improved distributional balance but produced little change in the estimated program effect.

The stability of the treatment-effect estimate is not evidence that the residual imbalance identified under the main-effects model was unimportant in general. Rather, it illustrates the role of balance diagnostics in improving the design independently of whether a particular refinement changes the substantive result. Approaches for addressing residual imbalance that persists after reasonable refinement of the weighting model are discussed in Section 5.

\section{Supporting Simulation Evidence}\label{sec:simulations}

In the empirical example, KS, AD, and CVM reached the same qualitative conclusions. Simulation studies were therefore used to examine whether that agreement generalizes when distributional discrepancies occur in different locations and forms, and to evaluate the calibration of alternative resampling procedures when weights are estimated from treatment assignment. Full simulation designs and results are provided in the Supplementary Material.

\subsection{Sensitivity to Different Forms of Distributional Imbalance}\label{sec:sim-sensitivity}

The simulations compared KS, AD, and CVM under centrally concentrated, tail-located, and diffuse distributional discrepancies. No test was uniformly most sensitive. With weights generated independently of group assignment, KS was most sensitive to the centrally concentrated discrepancy, AD was substantially more sensitive to the tail discrepancy, and AD and CVM were similarly sensitive to the diffuse discrepancy and both more sensitive than KS. These comparative sensitivity patterns were generally stable across different levels of weight variability, although modest anticonservatism appeared for all three tests at the most extreme weight variability examined.

Some findings changed under the estimated-weight designs examined. AD retained its pronounced advantage for tail discrepancies, and AD and CVM remained more sensitive than KS to diffuse discrepancies. For the centrally concentrated discrepancy, however, AD was more sensitive than KS under both estimated-weight procedures examined. Thus, although the relative sensitivity of the tests clearly depends on the form and location of the distributional discrepancy, the ordering observed under exogenously generated weights did not necessarily persist in the estimated-weight designs.

These findings support using KS, AD, and CVM as complementary diagnostics rather than selecting one as a universal measure of distributional balance. They also reinforce the value of examining weighted ECDFs, which show where discrepancies occur and provide context for differences among the test results.

\subsection{Resampling Procedures With Estimated Weights}\label{sec:sim-permutation}

A second set of simulations examined the calibration of resampling-based \textit{p}-values when weights were estimated from treatment assignment. Three procedures were considered: holding the estimated weights fixed while permuting treatment labels, re-estimating the propensity-score model after permutation, and generating new treatment assignments from an estimated assignment mechanism.

When the compared covariate was unrelated to treatment, fixed-weight permutation produced approximately nominal rejection rates across the treatment-assignment strengths examined. When the compared covariate was related to treatment assignment, fixed-weight permutation was instead conservative. Re-estimating the propensity-score model after unrestricted permutation could be substantially anticonservative. The distortion increased as the weighting regime generated under permutation diverged from that of the observed data and was not explained by weight variability alone. Assignment-mechanism resampling produced approximately nominal rejection rates when the propensity-score model was correctly specified, but could be miscalibrated when that model was misspecified.

The statistics were not affected equally. Under weight re-estimation, KS and CVM became increasingly anticonservative as the mismatch between observed and resampled weighting regimes increased, whereas AD remained substantially closer to nominal levels. This does not establish AD as generally calibrated under weight re-estimation; rather, it demonstrates that the behavior of the resulting \textit{p}-value depends jointly on the statistic and the procedure used to construct its reference distribution.

\section{Discussion}\label{sec:discussion}

The empirical example and simulations identify several considerations for researchers in the health professions using weighting methods in observational evaluations. Distributional diagnostics can reveal aspects of balance not captured by conventional summaries, different diagnostics may respond to different forms of residual imbalance, and resampling-based results depend in part on how their reference distributions are constructed. The practical implication is not that conventional diagnostics should be replaced, but that balance assessment should draw on complementary information and be used to improve the design before treatment effects are examined.

\subsection{Practical Guidance for Assessing and Addressing Distributional Imbalance}\label{sec:guidance}

The objective of balance assessment is to identify remaining differences between treatment groups, determine whether the weighting specification can reasonably be improved, and address important residual imbalance when it cannot. These decisions should be made without reference to the estimated treatment effect, preserving the separation between design and outcome analysis \citep{Rubin2001}.

\subsubsection{Assess and Interpret Balance}\label{sec:g-assess}

Balance should be assessed for all measured baseline covariates included in the weighting model. SMDs describe differences in location, while VRs provide additional information about dispersion for continuous covariates, and the distribution of the weights should be examined for extreme values or excessive variability \citep{AustinStuart2015}. For continuous and ordinal covariates with a sufficient number of distinct values, weighted ECDFs can show where treatment-group distributions differ, while KS, AD, and CVM provide complementary numerical assessments. For binary or categorical covariates with few levels, category proportions generally provide the relevant distributional information.

No individual diagnostic should serve as a pass-fail determination of balance. The empirical example demonstrates the limitation of using a cutpoint as a stopping rule, while the simulations show that no single distributional statistic is uniformly most sensitive to residual differences. Weighted ECDFs are particularly useful for placing the numerical results in context by showing where discrepancies occur \citep{lindengraphic2015}.

Resampling-based \textit{p}-values require an additional qualification. Their interpretation depends on the procedure used to construct the reference distribution and, when treatment-dependent weights are used, on how those weights are handled. The simulations also show that calibration can vary with propensity-model specification and the relationship between the tested covariate and treatment assignment. The resampling procedure should therefore be reported explicitly, and its \textit{p}-value interpreted as one component of the balance assessment rather than as evidence that a covariate is either balanced or imbalanced.

\subsubsection{Refine the Weighting Model and Reassess}\label{sec:g-refine}

When meaningful residual imbalance is identified, the weighting specification should be reconsidered. For propensity-score models, this may involve more flexible functional forms for continuous covariates, such as polynomial or spline terms, or the addition of interactions. New weights should then be estimated and balance reassessed for all measured covariates, not only those that prompted the refinement.

Alternative specifications should be compared according to the overall pattern and magnitude of residual imbalance rather than according to whether individual statistics cross prespecified thresholds. Refinement decisions should be based on balance without examining how alternative specifications affect the estimated treatment effect. The distribution of the resulting weights should also be reassessed because greater model flexibility may produce more variable or extreme weights.

\subsubsection{Address Remaining Imbalance and Report the Analysis}\label{sec:g-address}

When important residual imbalance remains after reasonable attempts to improve the weighting specification, affected covariates can additionally be included in the outcome model. Doubly robust estimators combine propensity-score adjustment with outcome regression and can provide consistent estimation when either the treatment-assignment model or the outcome model is correctly specified, subject to the assumptions of the estimator \citep{Bang2005,cattaneo2010,Linden2017}. Outcome adjustment should not, however, be viewed as a remedy for serious lack of overlap or comparability between treatment groups.

The balance-assessment process should be reported transparently. Reports should describe the weighting model and type of weights, including any trimming or truncation; present balance diagnostics for all measured covariates; describe how distributional balance was evaluated; and specify the procedure used to construct any resampling reference distribution. Weighting-model refinements, important remaining imbalances, and subsequent outcome-model adjustments should also be documented.

\subsection{Limitations}\label{sec:limitations}

The empirical example involved a single disease management program, one weighting method (stabilized inverse probability of treatment weights for the average treatment effect), and two continuous covariates selected to illustrate the problem. The patterns observed there, including the agreement and disagreement among conventional and distributional diagnostics, may differ for other programs, populations, covariates, and weighting approaches, such as overlap or entropy-balancing weights, and the example is intended to demonstrate the workflow rather than to establish how often such discrepancies arise in practice.

The simulations necessarily examined a finite set of data-generating mechanisms, discrepancy patterns, and weighting and resampling conditions. Their results are empirical operating characteristics under those conditions, not general theoretical results establishing the validity of the resampling procedures, and the calibration of the resampling procedures and the relative power of KS, AD, and CVM may differ under other designs, sample sizes, or propensity-model specifications. In particular, the conclusions concerning fixed-weight permutation, weight re-estimation, and assignment-mechanism resampling should be viewed as guidance about where caution is warranted rather than as proof of general behavior.

Finally, only the balance of measured covariates can be assessed. Even a weighted sample that is well balanced on every measured covariate, in terms of both summary statistics and complete distributions, provides no assurance about unmeasured confounders, and none of the diagnostics considered here addresses unmeasured confounding or other threats to causal inference.

\section{Conclusion}\label{sec:conclusion}

Covariate balance after weighting is best viewed as a multidimensional property rather than a pass-fail determination based on a single statistic. For researchers in the health professions, combining conventional balance measures with graphical and distributional diagnostics can identify residual differences, guide refinement of the weighting specification, and clarify what imbalance remains to be addressed before estimating treatment effects. The goal is not to find a statistic that declares balance, but to use complementary evidence to construct the most comparable treatment groups that the observed data reasonably permit.

\bibliographystyle{plainnat}
\bibliography{refs}

\clearpage
\begin{table}[p]
\centering
\caption{Balance Diagnostics for the Rx Score and Charlson Comorbidity Index (CCI) Score Before and After Weighting}
\label{tab:balance}
\small
\setlength{\tabcolsep}{6pt}
\begin{tabular}{llccc}
\toprule
 & & Unweighted & Main effects & Quadratic terms\\
\midrule
\multicolumn{5}{l}{\textit{Rx score}}\\
 & SMD & 0.191 & 0.046 & 0.041\\
 & VR & 1.077 & 0.807 & 1.096\\
 & KS $D$ & 0.1294 & 0.0706 & 0.0375\\
\addlinespace[0.6ex]
 & KS $p$ & $<$.001 & .007 & .300\\
 & CVM $p$ & $<$.001 & .004 & .214\\
 & AD $p$ & $<$.001 & .004 & .231\\
\addlinespace
\multicolumn{5}{l}{\textit{CCI score}}\\
 & SMD & 0.211 & 0.034 & 0.039\\
 & VR & 0.967 & 0.801 & 1.060\\
 & KS $D$ & 0.1229 & 0.0579 & 0.0326\\
\addlinespace[0.6ex]
 & KS $p$ & $<$.001 & .002 & .170\\
 & CVM $p$ & $<$.001 & .009 & .400\\
 & AD $p$ & $<$.001 & .003 & .326\\
\bottomrule
\end{tabular}

\vspace{1ex}
\begin{minipage}{0.95\textwidth}
\footnotesize
\textit{Note.} $N=5{,}907$ (766 participants, 5,141 non-participants). SMD = absolute standardized mean difference; VR = variance ratio (participants/non-participants); KS = Kolmogorov--Smirnov; CVM = Cram\'er--von Mises; AD = Anderson--Darling. Weighted columns use stabilized inverse probability of treatment weights from the main-effects model and from the model with quadratic terms for both covariates. Permutation \textit{p}-values are based on 2,000 permutations (weights held fixed in the weighted analyses); $p<.001$ is the smallest attainable value.
\end{minipage}
\end{table}

\clearpage
\begin{table}[p]
\centering
\caption{Estimated Pre--Post Difference in All-Cause Hospital Admissions per Patient}
\label{tab:effects}
\small
\setlength{\tabcolsep}{5pt}
\begin{tabular}{lcccc}
\toprule
 & \multicolumn{2}{c}{Mean change} & & \\
\cmidrule(lr){2-3}
Specification & Participants & Non-participants & Difference (\textit{SE}) & 95\% CI\\
\midrule
Unadjusted & $-$0.275 & $-$0.335 & 0.060 (0.047) & [$-$0.033, 0.153]\\
Main-effects weights & $-$0.229 & $-$0.341 & 0.112 (0.052) & [0.010, 0.215]\\
Quadratic-term weights & $-$0.218 & $-$0.342 & 0.124 (0.051) & [0.023, 0.224]\\
\bottomrule
\end{tabular}

\vspace{1ex}
\begin{minipage}{0.95\textwidth}
\footnotesize
\textit{Note.} $N=5{,}907$. Mean change is the intervention-period minus baseline difference in all-cause admissions per patient. Differences are participants minus non-participants, estimated by least squares with robust standard errors (weighted with stabilized weights where indicated). Positive differences indicate a smaller reduction in admissions among participants.
\end{minipage}
\end{table}

\clearpage

\clearpage

\end{document}